\documentclass[reprint,letterpaper,twocolumn,amssymb,nobibnotes,showpacs, aps, pra]{revtex4-1}
\usepackage{graphicx}
\usepackage{bm}

\begin{document}

\title{Magic Wavelength for the Hydrogen 1S-2S Transition}

\author{Akio Kawasaki}
\email{akiok@mit.edu}
\affiliation{Department of Physics, MIT-Harvard Center for Ultracold Atoms and Research Laboratory of Electronics, Massachusetts Institute of Technology, Cambridge, Massachusetts 02139, USA}


\begin{abstract}
The magic wavelength for an optical lattice for hydrogen atoms that cancels the lowest order AC Stark shift of the 1S-2S transition is calculated to be 513 nm. The magnitude of AC Stark shift $\Delta E=-1.19$ kHz/(10kW/cm$^2$) and the slope $d\Delta E/d\nu = -27.7$ Hz/(GHz $\cdot$ 10 kW/cm$^2$) at the magic wavelength suggests that a stable and narrow linewidth trapping laser is necessary to achieve a deep enough optical lattice to confine hydrogen atoms in a way that gives a small enough light shift for the precision spectroscopy of the 1S-2S transition.  
\end{abstract}

\maketitle
\section{Introduction}\label{Introduction}
Hydrogen spectroscopy is of fundamental interest to physicists, and has contributed to the development of quantum mechanics and quantum electrodynamics \cite{HAtom}. The spectroscopy has become more and more precise as new technologies have developed.  In particular, the precision of the spectroscopy of the 1S-2S transition has improved by ten orders of magnitude in the past century, and now the fractional frequency uncertainty is on the order of $10^{-15}$ \cite{RevModPhys.71.S242,PhysRevLett.107.203001}. This high precision contributes to the determination of several fundamental constants, such as Rydberg constant and proton radius \cite{RevModPhys.84.1527}, and is also used to set limits on the time variation of the fundamental constants \cite{PhysRevLett.92.230802} and the violation of Lorentz boost invariance \cite{PhysRevD.81.041701}.  It is also planned to test the CPT theorem by comparing the transition frequency of hydrogen and anti-hydrogen \cite{Nature.483.439,NatCommun.5.3089,PhysRevLett.108.113002}. 

So far, the precision spectroscopy of the hydrogen 1S-2S transition has been performed with a hydrogen atomic beam, because of the difficulty in trapping and cooling hydrogen \cite{PhysRevLett.107.203001}. Spectroscopy with an atomic beam cannot avoid the uncertainty due to the limited amount of interrogation time and the Doppler effect, and indeed the 2nd order Doppler effect is one of the major sources of the frequency uncertainty in Ref. \cite{PhysRevLett.107.203001}. The precision spectroscopy of other atomic species, on the other hand, is typically performed with trapped atoms or ions and takes advantage of the long interrogation time and the Lamb-Dicke regime confinement, which results in better relative uncertainty \cite{RevModPhys.87.637,PhysRevLett.114.230801}. For the hydrogen 1S-2S transition, one would also expect that spectroscopy with trapped atoms would improve the precision. 

To trap neutral atoms for precision spectroscopy, an optical lattice formed by a standing wave of laser light is typically used. The light-induced AC Stark shift becomes a trapping potential for atoms, but since the amount of the AC Stark shift is generally different for different states, the laser light also induces a frequency shift in optical transitions. At a special wavelength for the trapping light called the magic wavelength, the AC Stark shifts for the ground state and an excited state are the same, which nearly leads to the cancellation of the energy shift of the transition. The idea of the magic wavelength was first proposed for the strontium clock transition \cite{PhysRevLett.91.173005}, and is now widely used in state of the art optical transition atomic clocks \cite{RevModPhys.87.637}.  

In this paper, the magic wavelength for the 1S-2S transition of hydrogen is calculated, and the possibility of trapping hydrogen in an optical lattice of the magic wavelength is discussed. The AC Stark shift for the hydrogen ground state has been widely calculated. However, some \cite{PhysRevLett.61.2673,ApplPhysB.78.817,JPhysB.41.135602} are for the purpose of high intensity laser applications, and some others \cite{PhysRevA.18.1853} are calculations in a general situation.  To the best of my knowledge, there has never been a report comparing the AC Stark shift of the ground state with that of the 2S state for precision spectroscopy.  

In addition to the 1S-2S spectroscopy of hydrogen, optical trapping is particularly important for anti-hydrogen spectroscopy, where an intense atomic beam for spectroscopy cannot be generated. The state of the art anti-hydrogen trap for spectroscopy is a magnetic trap, and some measurements were performed for the ground state hyperfine transition \cite{Nature.483.439}, where the effect of the magnetic field is removed by subtracting two frequencies from transitions between different sub-levels.  With a magnetic field-free measurement in an optical lattice, the effect of the magnetic field is automatically removed, and the overall sequence to reduce magnetic field effects becomes simpler. 

\section{Calculation}
To calculate the trapping depth by an optical lattice for the ground state and an excited state, I calculate the AC polarizability of atoms in those states. Typically, such a calculation is performed with relativistic many-body perturbation theory
\cite{PhysRevLett.91.173005,PhysRevA.69.021403,PhysRevA.78.014502}, but for hydrogen, simple non-relativistic perturbation theory with the analytic solution of Schr\"{o}dinger equation can be used.  Some of the previous reports on the hydrogen ground state AC Stark shift have also used this simple method \cite{PhysRevLett.61.2673,ApplPhysB.78.817}. 

The two lowest order energy shifts of a state due to the oscillating electric field are given as
\begin{equation}
\Delta E=-\frac{1}{4}\alpha({\bf e},\omega) {\cal E} ^{2}-\frac{1}{64}\gamma({\bf e},\omega) {\cal E} ^{4}-\cdots,
\end{equation}
where ${\cal E}$, $\alpha$, $\gamma$, ${\bf e}$ and $\omega$ are the amplitude of the electric field, polarizability, hyperpolarizability, the polarization of the light and the light frequency \cite{PhysRevLett.91.173005}. The largest contribution to the polarizability is from electric dipole (E1) transitions, and electric quadrupole (E2) and magnetic dipole (M1) transitions have the second largest contribution:
\begin{equation}
\alpha({\bf e},\omega)=\alpha_{E1}({\bf e},\omega)+\alpha_{M1}({\bf e},\omega)+\alpha_{E2}({\bf e},\omega)+\cdots
\end{equation}
Second-order perturbation theory gives $\alpha_{E1}({\bf e},\omega)$ for a state $|n\rangle$ as
\begin{equation}\label{ACStarkShift1st}
\alpha_{E1}({\bf e},\omega)=\frac{2}{\hbar}\sum_k \frac{\omega_{kn}|\langle k|{\bf d \cdot e}|n \rangle|^2}{\omega_{kn}^2-\omega^2},
\end{equation}
where $\hbar \omega_{kn}$ is the energy difference of the state $|k\rangle$ and $|n\rangle$ and ${\bf d}$ is the operator for dipole moment. For  hydrogen, the exact energy levels and wave functions can be found as the solutions of the Schr\"{o}dinger equation in a nonrelativistic treatment, while relativistic correction, given by the difference between the solution of the Schr\"{o}dinger equation and the solution of the Dirac equation, can be regarded as a higher order correction. I first calculate the lowest order shift, and then estimate the corrections due to higher-order terms.  

The matrix element is separated into the angular component and the radial component. The angular component involves the Clebsch-Gordan coefficients, and is generally magnetic sub-level dependent. In the case of the nS-n'P transitions in hydrogen, which are the only allowed E1 transitions from nS states, the transitions between $F=0$ component of a nS state and $F=1$ component of a n'P state will have polarization independent Clebsch-Gordan coefficients. For a transition between a $^2$S$_{1/2}$ $F=0$ state and a $^2$P$_{1/2}$ $F=1$ state, the coefficient is  $1/\sqrt{3}$, while for a transition between a $^2$S$_{1/2}$ $F=0$ state and a $^2$P$_{3/2}$ $F=1$ state, the coefficient is $\sqrt{2/3}$. Since the polarization independent AC Stark shift by the trapping light is desired, I assume that the spectroscopy is performed between $F=0$ sublevels of the 1S and 2S states and regard $1/\sqrt{3}$ and $\sqrt{2/3}$ as the angular components of the matrix elements respectively.  

The general form of the radial wave function of the hydrogen atom is
\begin{equation}\label{Rnl}
R_{n,l}(\rho)=\sqrt{\frac{4(n-l-1)!}{n^4 [(n+l)!]}}\left( \frac{Z}{a_0}\right)^{3/2} \rho^l e^{-\rho/2} L^{2l+1}_{n-l-1}(\rho) 
\end{equation}
where $a_0$ is the Bohr radius and
\begin{equation}\label{rho}
\rho=\frac{2Zr}{na_0}.   
\end{equation}
With Eqs. \ref{Rnl} and \ref{rho}, the radial components become
\begin{widetext}
\begin{eqnarray}
\langle R_{k,1}| r | R_{1,0} \rangle &=& \frac{k^2}{4} \frac{(k+1)!}{\sqrt{(k+1)k(k-1)}}
  \sum^{k-2}_{m=0} \frac{(-1)^m (m+4)}{(k-m-2)!m!}\left( \frac{2}{k+1} \right)^{m+5} \frac{a_0}{Z}  \label{MatrixRadial1S}\\
\langle R_{k,1}| r | R_{2,0} \rangle &=& \frac{k^2}{8\sqrt{2}} \frac{(k+1)!}{\sqrt{(k+1)k(k-1)}}
\sum^{k-2}_{m=0} \frac{(-1)^{m+1}(m+4)}{(k-m-2)!m!}\frac{4k+km-2}{k+2} \left( \frac{4}{k+2} \right)^{m+5} 
\frac{a_0}{Z} \label{MatrixRadial2S}
\end{eqnarray}
Combining the angular components, Eqs. \ref{ACStarkShift1st}, \ref{MatrixRadial1S} and \ref{MatrixRadial2S}, and setting $Z=1$, I obtain
\begin{eqnarray}
\alpha^{1S}_{E1}({\bf e}_z, \omega) &=& \frac{e^2 a_0^2}{16 \hbar ^2}mc^2 \alpha^2 \sum^{\infty}_{n=2}\frac{1}{\omega_{n1}^2-\omega^2}\frac{(n+1)!}{n}\left[ \sum^{n-2}_{m=0} \frac{(-1)^m  (m+4)}{(n-m-2)! m!}\left( \frac{2}{n+1} \right)^{m+5} \right]^2 \\
\alpha^{2S}_{E1}({\bf e}_z, \omega) &=& \frac{e^2 a_0^2}{128 \hbar ^2}mc^2 \alpha^2 \sum^{\infty}_{n=3}\frac{1}{\omega_{n2}^2-\omega^2} \frac{n-2}{n+2} \frac{n[(n+1)!]^2}{n^2-1} \left[ \sum^{n-2}_{m=0} \frac{(-1)^{m+1} (m+4)}{(n-m-2)! m!}\left( \frac{4}{n+2} \right)^{m+5} (4n+nm-2) \right]^2, 
\end{eqnarray}
\end{widetext}
where $\alpha$ is the fine structure constant. The summation over $n$ does not have any simpler analytical form and therefore can be calculated numerically with a large enough upper limit $n_{\rm max}$.  

\begin{figure}[!t]
	\begin{center}
 \includegraphics[width=1.0\columnwidth]{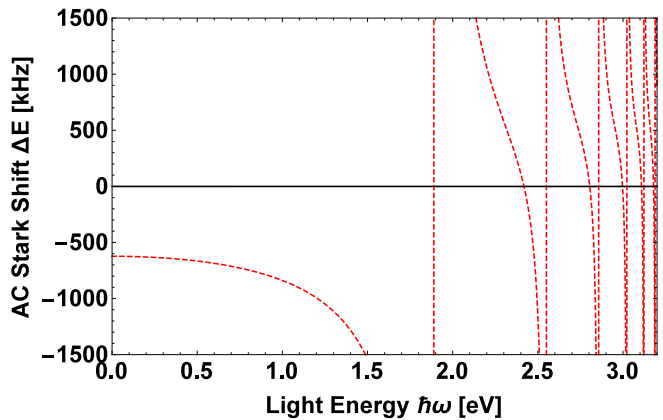}
 \caption{
(Color online)
AC Stark shift for the 1S (black solid line) and the 2S (red dashed line) state by visible light. Intensity is 10kW/cm$^2$  and $n_{\rm max}=100$.} 
 \label{BBPlot}
 \end{center}
\end{figure}

Figure \ref{BBPlot} shows the calculated AC Stark shift with $n_{\rm max}=100$ for visible light. The 1S state has an almost constant AC Stark shift in this region, while the 2S state shift changes considerably. This is because the minimum transition energy for the 1S state is 10.2 eV, and visible light is far red-detuned for all transitions from the 1S state.  This, in turns, means when the AC Stark shift for the 1S state changes significantly due to the transition between the 1S and nP states, the shift for the 2S state is more or less constant and positive.  Since a negative AC Stark shift is required for the one dimensional optical lattice, the visible light region is of interest.  

\begin{figure}[!t]
	\begin{center}
 \includegraphics[width=1.0\columnwidth]{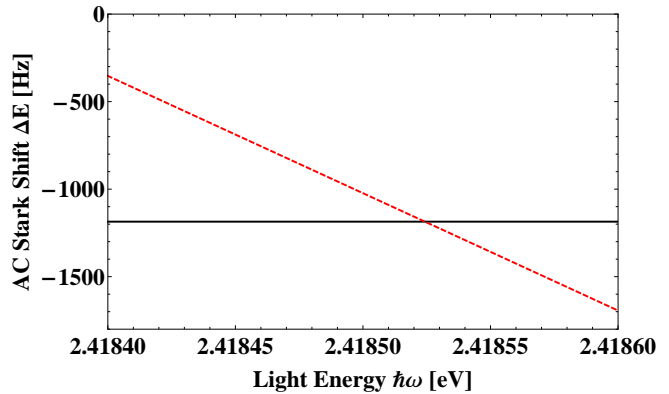}
 \caption{
(Color online)
AC Stark shift for the 1S (black solid line) and the 2S (red dashed line) state around 2.4185 eV (512.64nm) by 10kW/cm$^2$ light. $n_{\rm max}=3000$.} 
 \label{ZoomPlot}
 \end{center}
\end{figure}

Figure \ref{BBPlot} suggests that there is a point where $\alpha^{1S}_{E1}=\alpha^{2S}_{E1}$ around $\hbar \omega \simeq 2.4$ eV, and Fig. \ref{ZoomPlot} is the magnified plot for this region. The crossing point is at 2.4185 eV, which is 512.64 nm in the units of wavelength. This is the lowest energy magic wavelength for the hydrogen 1S-2S transition. Based on the fact that there are shorter wavelength transitions for the 2S state, there are more magic wavelengths for the hydrogen 1S-2S transition, such as 2.806 eV (441.8 nm) and 2.997 eV (413.7 nm).  However, 2.4185 eV is the best magic wavelength in the sense that the slope $d\Delta E/ d \omega$ is the smallest.  Thus, I will concentrate on the magic wavelength of 512.64 nm.

\section{Effects of higher order terms}
Next, I estimate the contribution by higher order terms. The effects of E2 and M1 transitions are calculated by perturbation theory.  The Hamiltonian for the E2 transition is
\begin{equation}
H_{E2}= \sum_{i,j} \frac{e}{2} \left( r_i r_j - \frac{r^2}{3} \delta_{ij}\right) \partial_i E_j,
\end{equation}
With a linearly polarized plane wave $ \mbox{\boldmath $E$}=E_0\mbox{\boldmath $e$}e^{i k \mbox{\boldmath $n$} \cdot \mbox{\boldmath $ x$} -\omega t}$, the polarizability by E2 transition is given as
\begin{equation}
\alpha_{E2}(\omega) =\frac{2}{\hbar}\frac{e^2\omega^2}{4c^2}\sum_{i,j,k}\frac{\omega_{kn}|\langle k | \left( n_i e_j - \frac{r^2}{3} \delta_{ij}\right) | n \rangle|^2}{\omega_{kn}^2-\omega^2},
\end{equation}
The polarizability of the 1S state due to the 3D state is, for instance, 
\begin{equation}
\alpha^{1S-3D}_{E2}(\omega)=\frac{3^7}{2^{14}}\frac{e^2\omega^2}{\hbar c^2}\left( \frac{a_0}{Z}\right)^4\frac{\omega_{31}}{\omega_{31}^2-\omega^2},
\end{equation}
where the angular component takes a value between 0 and 4/45 depending on the direction of $\mbox{\boldmath $e$}$ and $\mbox{\boldmath $n$}$, and 4/45 was used to set an upper limit. The polarization of the 1S state by the 3P state is
\begin{equation}
\alpha^{1S-3P}_{E1}(\omega)=\frac{3^6}{2^{12}}\frac{e^2}{\hbar}\left( \frac{a_0}{Z}\right)^2\frac{\omega_{31}}{\omega_{31}^2-\omega^2},
\end{equation}
and the difference is the factor of $\omega^2a_0^2/c^2$, except for the numerical prefactor.  With visible light and an atom, this factor is negligibly small and thus $\alpha_{E2}$ is expected to be negligible compared to $\alpha_{E1}$.  At 512.64 nm, the ratio becomes $2.37 \times 10^{-7}$. Because this factor of $\omega^2a_0^2/c^2$ is dominant in the ratio $\alpha^{1S-3D}_{E2}/\alpha^{1S-3P}_{E1}$, the ratio of the polarizability due to other excited states $\alpha^{1S-nD}_{E2}/\alpha^{1S-nP}_{E1}$ is also expected to be small. Thus total polarizability due to E2 transitions $\alpha_{E2}(\omega)$ is negligible compared to $\alpha_{E1}(\omega)$.

The Hamiltonian for the M1 transitions is 
\begin{equation}
H_{M1}=\frac{\mu_B}{\hbar}\left(\mbox{\boldmath $L$}+g\mbox{\boldmath $S$}\right)\cdot \mbox{\boldmath $B$}
\end{equation}
and this gives 
\begin{equation}
\alpha_{M1}(\omega) =\frac{2}{\hbar^3} \mu_B^2
\sum_{k}\frac{\omega_{kn}|\langle k | \left(\mbox{\boldmath $L$}+g\mbox{\boldmath $S$}\right)\cdot \mbox{\boldmath $e$} | n \rangle|^2}{\omega_{kn}^2-\omega^2}
\end{equation}
for a linearly polarized plane wave $ \mbox{\boldmath $B$}=(E_0/c)\mbox{\boldmath $e$}e^{i k\mbox{\boldmath $ n$} \cdot \mbox{\boldmath $ x$} -\omega t}$. This is significantly smaller than $\alpha_{E1}(\omega)$ due to two factors. The factor $\mu_b^2 $ is $\alpha^2$ smaller than $e^2 a_0^2$. The matrix element suggests that there is no change in electronic structure, and thus the only allowed M1 transitions for hydrogen are hyperfine transitions. This results in $\omega_{kn}$ smaller by factor of $10^5$ compared to the E1 transition case. Thus, $\alpha_{M1}(\omega)$ is negligible compared to $\alpha_{E1}(\omega)$. 

Hyperpolarizability $\gamma({\bf e},\omega)$ is induced by higher order perturbative interactions between an atom and two photons. Given that the electric field of the light is much smaller than the internal field in the atoms, as shown by the six or more orders of magnitude smaller AC Stark shift for typical trapping depth than the atomic energy, the perturbative expansion is a good approximation. In this case, $\gamma({\bf e},\omega)$ is significantly smaller than $\alpha({\bf e},\omega)$, unless there is a two photon resonance.  Since neither the 1S state or the 2S state has a transition of 4.8 eV, no two photon transition is expected to give a significant contribution to $\gamma({\bf e},\omega)$.  

The relativistic correction is divided into a correction of the energy and the correction of the wave function. These are expected to be small, as the velocity of the electron is $c\alpha$, and is thus not relativistic. Based on the Dirac equation for the hydrogen atom, I can estimate the effect with analytic solutions. The energy level correction includes a factor of $\omega_{kn}/(\omega_{kn}^2-\omega^2)$.  For example, the relativistic correction to the 1S$_{1/2}$ state and the 2P$_{3/2}$ state are -43.8 GHz and -2.74 GHz, which gives the change of 41.1 GHz in $\omega_{21}$.  This shift in $\omega_{21}$ gives the relative change of $\omega_{21}/(\omega_{21}^2-\omega^2)$ of $1.86 \times 10^{-5}$ at the wavelength of 512.64 nm. 

The relativistic correction in the wave function slightly changes the matrix element. Defining the difference between the relativistic wave function $|\Psi^{rel}_{1S} \rangle$ and the nonrelativistic wave function $|\Psi^{nonrel}_{1S} \rangle$ as $|\Psi^{\delta}_{1S} \rangle$, I obtain
\begin{eqnarray}
\langle \Psi^{rel}_{2S} |r|\Psi^{rel}_{1S} \rangle &\simeq& 
\langle \Psi^{nonrel}_{2S} |r|\Psi^{nonrel}_{1S} \rangle \nonumber \\
&&+ \langle \Psi^{\delta}_{2S} |r|\Psi^{nonrel}_{1S} \rangle
+ \langle \Psi^{nonrel}_{2S} |r|\Psi^{\delta}_{1S} \rangle \nonumber \\
&=& \langle \Psi^{nonrel}_{2S} |r|\Psi^{rel}_{1S} \rangle
+ \langle \Psi^{\delta}_{2S} |r|\Psi^{nonrel}_{1S} \rangle,
\end{eqnarray}
assuming the correction is small. We calculate $\langle \Psi^{nonrel}_{2S} |r|\Psi^{rel}_{1S} \rangle$ to estimate $\langle \Psi^{nonrel}_{2S} |r|\Psi^{\delta}_{1S} \rangle$. As
\begin{eqnarray}
|\Psi^{rel}_{1S} \rangle &=& Y^0_0(\theta,\phi)\sqrt{2\Gamma(1+2\gamma})  \nonumber \\
 & &\ \ \ \times \left( \frac{Z}{a_0} \right)^{3/2} \left( \frac{Z}{a_0} r\right) ^\gamma e^{-Zr/a_0},
\end{eqnarray}
where $\gamma=\sqrt{1-\alpha^2}$, the matrix elements are
\begin{eqnarray}
\langle \Psi^{nonrel}_{2S} |r|\Psi^{rel}_{1S} \rangle 
&=&\frac{1}{2} \sqrt{\frac{\Gamma(1+2\gamma)}{3}} \left(\frac{2}{3}\right)^{4+\gamma} \Gamma \left(4+\gamma \right) \frac{a_0}{Z} \nonumber \\
&\simeq& 1.29020 \frac{a_0}{Z} \\
\langle \Psi^{nonrel}_{2S} |r|\Psi^{nonrel}_{1S} \rangle 
&=& \sqrt{\frac{2}{3}}\frac{128}{81} \frac{a_0}{Z} \nonumber \\
&\simeq& 1.29026 \frac{a_0}{Z}
\end{eqnarray}
The difference is $\langle \Psi^{nonrel}_{2S} |r|\Psi^{\delta}_{1S} \rangle=5.4 \times 10^{-5}\langle \Psi^{nonrel}_{2S} |r|\Psi^{nonrel}_{1S} \rangle$. $\langle \Psi^{\delta}_{2S} |r|\Psi^{nonrel}_{1S} \rangle$ is expected to be on the same order of magnitude, and therefore the overall relativistic correction to the wave function is negligible. 

All other effects on the energy levels, such as the Lamb shift, the finite nucleus size effect and the hyperfine splitting are around 1 GHz or less, which corresponds to a fractional amount of $10^{-6}$ or less. Thus, the overall higher order correction is at most on the order of $10^{-4}$, and therefore the number 512.64 nm is reliable up to the three digit precision.  

\begin{table}[tb]
 \caption{Scattering rate by lattice light for relevant transitions} 
\begin{tabular}{cccc}
 \hline  \hline 
Transition & $\Gamma$ [s$^{-1}$]	&$I_{\rm sat}$ [mW/cm$^2$] & R [s$^{-1}$]\\
 \hline  
1S $\rightarrow$ 2P 	&$6.26 \times 10^8$	&7256	&$7.9 \times 10^{-3}$	\\
2S $\rightarrow$ 3P 	&$2.24 \times 10^7$	&1.65	&$1.4 \times 10^{-3}$	\\
2S $\rightarrow$ 4P 	&$9.67 \times 10^6$	&1.75	&$1.6 \times 10^{-3}$	 \\
   \hline  \hline 
\end{tabular}
\label{ScatterRate}
\end{table}

\section{Implementation of the magic wavelength optical lattice}
Figure \ref{ZoomPlot} shows that the AC Stark shift at the magic wavelength is -1.19 kHz/(10kW/cm$^2$).  This is 50 times smaller than the alkali earth-like atoms like strontium or ytterbium \cite{LaserPhys.15.1040}.  In addition, the recoil energy for hydrogen is 72 $\mu$K, due to its small mass. These two factors require very intense light for hydrogen trapping.  In order to obtain a trap depth for hydrogen of 300$E_r$ trapping depth, which is typical for the state of the art optical lattice clocks \cite{Science.341.1215,Nature.506.71}, $3.8 \times 10^6$ kW/cm$^2$ intensity is required for hydrogen trapping. This is barely achievable by focusing 1 W light injected to an optical cavity of finesse 3,000$\pi$ that has a beam waist diameter of 10 $\mu$m.  

The slope of the polarizability is $d\Delta \alpha /d\nu=-27.7$ Hz/GHz for 10kW/cm$^2$ light. The 300$E_r$ lattice gives a -10.5 MHz/GHz shift.  Compared to the ytterbium magic wavelength trap that gives a slope of 11(1) Hz/GHz for 500$E_r$ lattice \cite{PhysRevLett.100.103002}, this is six orders of magnitude larger. In order to suppress the frequency uncertainty of the 1S-2S transition due to the light shift to 1Hz or lower, the linewidth of the trapping light should be 100 Hz or less, and the magic wavelength needs to be determined with a similar accuracy.  

Given the high intensity of the lattice, the loss from the lattice due to the scattering of the lattice light becomes a concern.  Table \ref{ScatterRate} summarizes the linewidth $\Gamma$, the saturation intensity $I_{\rm sat}$, and $R$, the scattering rate at 512.64 nm for the three closest transitions involving the 1S and 2S states.  The rate is significantly smaller than 1 s$^{-1}$ and therefore the loss due to the scattering is not a concern.  However, it should be noted that the small mass of hydrogen and the optical cavity to enhance the power might complicate suppression of the heating due to the lattice intensity fluctuations \cite{PhysRevA.56.R1095}.  

Another practical concern is cooling hydrogen to a temperature cold enough to trap into the lattice. Hydrogen was first trapped in a magnetic trap with buffer gas cooling and then evaporatively cooled down to 50 $\mu$K to achieve a Bose-Einstein condensate \cite{PhysRevLett.81.3811}.  This is cold enough for atoms to be loaded into the optical lattice, but this method does not work for anti-hydrogen, as the number of atoms that can be trapped is only a few in each cycle of the experiment. A cooling scheme with the Lyman-$\alpha$ transition was recently proposed \cite{JPhysB.46.025302}, and the predicted achievable temperature was 20 mK. It would be possible to trap a few anti-hydrogen atoms in the optical lattice, but a more sophisticated way to cool anti-hydrogen would be necessary. 

\section{Conclusion}
The magic wavelength for the hydrogen 1S-2S transition is calculated using the solutions of the Schr\"{o}dinger equation.  The wavelength is estimated to be 513 nm; the trap depth is -1.19 kHz per 10kW/cm$^2$ intensity; and the slope of the transition frequency shift is -10.5 MHz/GHz for a 300 $E_r$ lattice depth.  These results imply that a sophisticated system is required to implement an optical lattice for hydrogen, such as power enhancement by an optical cavity and a narrow linewidth laser. However, it should be possible to trap hydrogen atoms in an optical lattice for the 1S-2S transition spectroscopy.  

\begin{acknowledgements}
The author thanks Boris Braverman for a helpful discussion. This work is supported by DARPA and NSF.  
\end{acknowledgements}

\bibliographystyle{apsrev4-1}
\bibliography{HMagicWavelengthPRA}

\end{document}